\documentclass[twocolumn]{aastex61}
\pdfoutput=1 
\usepackage{amsmath,amstext}
\usepackage[T1]{fontenc}
\usepackage{apjfonts} 
\usepackage[figure,figure*]{hypcap}

\shorttitle{Return Currents in NS Atmospheres}
\shortauthors{Baub\"ock et al.}

\begin{document}

\title{Atmospheric Structure and Radiation Pattern for Neutron-Star Polar Caps Heated by Magnetospheric Return Currents}

\author{Michi Baub\"ock}
\affil{Max Planck Institut f\"ur Extraterrestrische Physik,
Gie{\ss}enbachstr.\ 1, D-85737 Garching, Germany}

\author{Dimitrios Psaltis}
\affil{Astronomy Department,
University of Arizona,
933 N.\ Cherry Ave.,
Tucson, AZ 85721, USA}

\author{Feryal \"Ozel}
\affil{Astronomy Department,
University of Arizona,
933 N.\ Cherry Ave.,
Tucson, AZ 85721, USA}


\begin{abstract}
The Neutron-star Interior Composition ExploreR (NICER) is collecting data to measure the radii of neutron stars by observing the pulsed emission from their surfaces. The primary targets are isolated, rotation-powered pulsars, in which the surface polar caps are heated by bombardment from magnetospheric currents of electrons and positrons.  We investigate various stopping mechanisms for the beams of particles that bombard the atmosphere and calculate the heat deposition, the atmospheric temperature profiles, and the energy spectra and beaming of the emerging radiation. We find that low-energy particles with $\gamma \sim 2 - 10$ deposit most of their energy in the upper regions of the atmosphere, at low optical depth, resulting in beaming patterns that are substantially different than those of deep-heated, radiative equilibrium models. Only particles with energies $\gamma \gtrsim 50$ penetrate to high optical depths and fulfill the conditions necessary for a deep-heating approximation. We discuss the implications of our work for modeling the pulse profiles from rotation-powered pulsars and for the inference of their radii with NICER observations.
\end{abstract}


\section{Introduction}
X-ray emission from neutron-star surfaces provides a unique opportunity to explore the extreme physics of their interiors and surroundings.  Precise measurements of this emission can help constrain open questions  about their surface properties, magnetic field configuration, and interior composition~(see, e.g., \citealt{oze13} for a review).

With the data it is collecting, the NICER mission \citep{gen16}  will enable measurements of X-ray pulse profiles to unprecedented accuracy.  These pulse profiles arise from temperature inhomogeneities on the stellar surface, usually in the form of a small region of higher temperature than the surroundings (a ``hotspot''). Of particular interest to this mission are isolated rotation-powered pulsars, on which an area near the magnetic pole is heated by the return currents that flow in the pulsar magnetospheres. Measuring the effects of gravitational self-lensing on the pulse profiles from such sources is expected to lead to measurements of their masses and radii~\citep{zav98,bog07,bog08,bog13,oze16b}. 

During the last decades, there has been substantial effort to calculate the effects of gravitational lensing on the pulse profiles and assess the prospect of using such profiles to measure neutron-star masses and radii~\citep{pec83,mil98,wei01,pou06,cad07,mor07,lo13,psa14b,psa14a}. One of the key results of these studies has been that accurate interpretation of the pulse profiles relies on a good understanding of the properties of the stellar atmosphere and, in particular, on the dependence of the emitted spectrum of radiation on the angle from the normal to the stellar surface. If the angular dependence of emission, which we will refer to hereafter as the beaming of the emerging radiation, is strongly peaked around the normal to the surface, high amplitudes of pulsations can be observed even from very compact neutron stars. In other words, the inferred compactness of the neutron star is highly correlated with the beaming of the radiation emerging from its surface.

Previous models of pulse profiles from rotation-powered pulsars have made the simple assumption that the surface emission spectrum is that of an isotropic blackbody or used the spectrum and beaming of an atmosphere in radiative equilibrium (see, e.g., \citealt{zav98}, \citealt{bog07}, \citealt{gui11}, \citealt{bog13}). The latter assumption is only valid if the magnetospheric particles that heat the polar caps deposit their energy at layers much deeper than the photosphere. In this deep heating regime, the atmosphere remains in radiative equilibrium and the resulting emission is indistinguishable from the thermal spectrum of an isolated, cooling neutron star of the same temperature. If, on the other hand, the particles impinging on the atmosphere deposit their energy near the photospheric layers, the resulting temperature profile and radiation spectrum and beaming will be very different from that of a cooling atmosphere.

In the context of the isolated, rotation-powered pulsars, which are the prime targets for NICER, the energetics of the magnetospheric particles have been studied primarily in order to understand the emission of high-energy radiation. \citet{rud75} constructed a model in which electron-positron pairs are formed in a magnetospheric gap above the polar cap due to the large potential difference caused by the outflow of electrons along magnetic field lines. In this model, the electrons formed in these pairs flow back towards the surface, while positrons escape to infinity along magnetic field lines. In subsequent years, numerous models have investigated the location and mechanism of the pair formation (e.g., \citealt{rud75}, \citealt{aro81}, \citealt{aro83}, \citealt{che86}, \citealt{har01}, \citealt{har02}, \citealt{bai10a}, \citealt{bai10b}, \citealt{phi2015a}, \citealt{phi2015b}, \citealt{che2017}, \citealt{par2017}, \citealt{phi2018}, \citealt{bra18}). These models have since been compared to various observational signatures, with the most recent being the detailed $\gamma$-ray spectra and pulse profiles  of millisecond pulsars obtained since the launch of the {\em Fermi\/}  satellite~(see, e.g., \citealt{bai10a}, \citealt{pie15}, \citealt{kal18},  and \citealt{har16} for a review). There has been, however, no effort so far to calculate the temperature profile of the bombarded neutron-star atmosphere and the resulting spectrum and beaming of the surface emission.

The physics of radiative cooling in neutron-star atmospheres is well understood, with the most severe uncertainties in the models arising from the unknown elemental abundances in the atmospheres. In the case of deep-heated, radiative equilibrium atmospheres, numerous calculations exist for non-magnetic~\citep[see, e.g.,][]{lon86, rom87, zav96,raj96,bog07,mad04,maj05,sul12,haa12}, magnetic~\citep[see, e.g.,][]{shi92,mil92,pot03,ho08,pot14}, and strongly magnetic neutron stars~\cite[see, e.g,][]{oze01,ho01,oze03a,ho03,van06}. In the case of the heated polar caps of rotation-power pulsars, which are the primary NICER targets, the additional physical process of the stopping of the beam of magnetospheric particles in the atmosphere and the differential heating of its various layers need to be incorporated and understood.

In this paper, we calculate the deposition of the energy of magnetospheric particles on the surface layers of a neutron-star polar cap. In \S2, we explore various stopping mechanisms and identify those that dominate the energy deposition rates. In \S3-5, we then develop a simple model for the atmosphere that allows us to calculate its temperature profile and, in \S6, the beaming and spectrum of the emerging radiation. 

We find that the structure of the atmosphere depends strongly on the energy of the particles in the return current. Relatively low-energy electrons and positrons ($\gamma\lesssim$10) deposit most of their energy at low optical depths. This shallow energy deposition leads to temperature inversions in the atmosphere and beaming patterns of the emerging radiation that are substantially different from those of radiative equilibrium models. High energy particles, on the other hand, have a much larger stopping depth and therefore deposit their energy at high optical depths. In this case, the temperature profile and the beaming pattern of radiation approach those of radiative equilibrium atmospheres. In \S7, we conclude with a discussion of the effect of the details of the energy deposition on the shapes of the pulse profiles from isolated X-ray pulsars and the inference of their masses and radii with NICER.

\section{Charged Particle Energy Losses in a Plasma}\label{section:Motivation}
Energetic electrons and positrons can interact with the particles of a plasma in several different ways. They can scatter directly off both the electrons and the ions in the plasma, or they can excite plasma waves. Additionally, particles can lose energy via the emission of Bremsstrahlung radiation as they pass near ions in the plasma. In this section, we will discuss the dominant interaction mechanisms and calculate the energy loss rate of both electrons and positrons traveling through a hydrogen plasma.  

The return current electrons and positrons we consider here are moderately to highly relativistic ($\gamma \sim 2$ to $\gamma \sim 500$). The electrons in the neutron-star atmosphere, in contrast, have energies near 1~keV, corresponding to $\gamma \sim 1.004$. Therefore, we can make the assumption that the plasma electrons and ions are essentially stationary compared to the return current particles. 

The first channel for energy loss is via binary interactions with the free electrons (M\o ller scattering or Bhabha scattering for electrons and positrons, respectively) and the ions in the plasma. Since the cross-section for the scattering interaction is inversely proportional to the mass of the target, we neglect the electron-ion and positron-ion scattering terms and calculate only the scattering with the plasma electrons. 

In addition to direct scattering, the relativistic particles excite collective plasma modes (Langmuir waves) and thereby transmit additional energy to the atmosphere. Direct scattering and excitation of Langmuir waves dominate the energy loss rate for particles passing through a neutron star atmosphere. Here, we calculate the combined energy loss rate from these two mechanisms. 

\citet{sol08} derive the energy loss per unit path length of electrons traveling through a plasma due to both binary interactions and collective modes: 
\begin{multline}
-\left(\frac{dE}{dz}\right)_{\rm e} = \frac{2\pi n_e e^4}{m_e \beta^2 c^2} \bigg[\ln\left(\frac{E^2}{\hbar^2 \omega_p^2} \frac{\gamma + 1}{2 \gamma^2}\right) + 1 + \frac{1}{8} \left(\frac{\gamma - 1}{\gamma}\right)^2\\
 - \left(\frac{2 \gamma - 1}{\gamma^2}\right)\ln 2 \bigg],
\label{eq:ee}
\end{multline}
where $n_e$ is the electron density in the plasma, $e$ is the electron charge, $m_e$ is the electron mass, $\beta$ and $\gamma$ are the usual relativistic factors, $c$ is the speed of light, $\hbar$ is the reduced Planck constant, and $\omega_p = \sqrt{4\pi n_e e^2/m_e}$ is the plasma frequency. The energy $E$ is related to the relativistic factor $\gamma$ through
\begin{equation}
E = (\gamma - 1) m_e c^2.
\label{eq:E_def}
\end{equation}

Positrons passing through a neutron-star atmosphere similarly lose energy via direct collisions with the plasma electrons and by exciting Langmuir waves. In this case, the cross-section of interaction is governed by the Bhabha equation rather than the M\o ller cross-section \citep{bha36}. \citet{roh54} derive the total energy loss per unit length of positrons traveling through a non-ionized medium,
\begin{equation}
-\left(\frac{dE}{dz}\right)_{\rm p} = \frac{2\pi n_e e^4}{m_e \beta^2 c^2} \left[\ln\left(\frac{E^2}{I^2} \frac{\gamma + 1}{2}\right) + f^+(\gamma)\right],
\label{eq:rc54}
\end{equation}
where $I$ is the average ionization energy of the medium, and $f^+$ is a function of $\gamma$, 
\begin{equation}
f^+(\gamma) = 2 \ln(2) - \frac{\beta^2}{12}\left[23 + \frac{14}{\gamma + 1} + \frac{10}{(\gamma + 1)^2} + \frac{4}{(\gamma + 1)^3}\right].
\label{eq:fplus}
\end{equation}

Following \citet{ICRU37}, we add a density-effect correction $-\delta$ to the term for $f^+$. In the limit $\gamma \rightarrow \infty$, the density-effect correction approaches
\begin{equation}
\delta \rightarrow \ln\left[\frac{(\hbar \omega_p)^2 \gamma^2}{I^2} \right] - 1.
\label{eq:delta}
\end{equation}
In this high-energy limit, the interaction time between a positron and a bound electron is very short and the equation for energy loss is the same for a non-ionized and an ionized medium \citep{sol08}. Therefore, we can include the expression for $\delta$ from equation (\ref{eq:delta}) in equation (\ref{eq:rc54}) to find the energy loss due to binary collisions of a positron passing through a plasma, i.e., 
\begin{multline}
-\left(\frac{dE}{dz}\right)_{\rm p} = \frac{2\pi n_e e^4}{m_e \beta^2 c^2} \bigg[\ln\left(\frac{E^2}{\hbar^2\omega_p^2}\frac{\gamma + 1}{2\gamma^2}\right) + 1 + 2\ln(2) - \\
\frac{\beta^2}{12}\left(23 + \frac{14}{\gamma + 1} + \frac{10}{\left(\gamma + 1\right)^2} + \frac{4}{\left(\gamma + 1\right)^3}\right)\bigg] 
\label{eq:pe}
\end{multline}                              

Lastly, electrons and positrons passing near ions lose energy via the emission of Bremsstrahlung radiation. The rate of energy loss due to Bremsstrahlung is proportional to the energy of the particle. While this mechanism can dominate for extremely relativistic electrons, for the particle energies we consider here, the energy lost due to Bremsstrahlung is several orders of magnitude smaller than the processes discussed above. We, therefore, neglect the contribution of the Bremsstrahlung radiation. 

\begin{figure}
\includegraphics[width=3.5in]{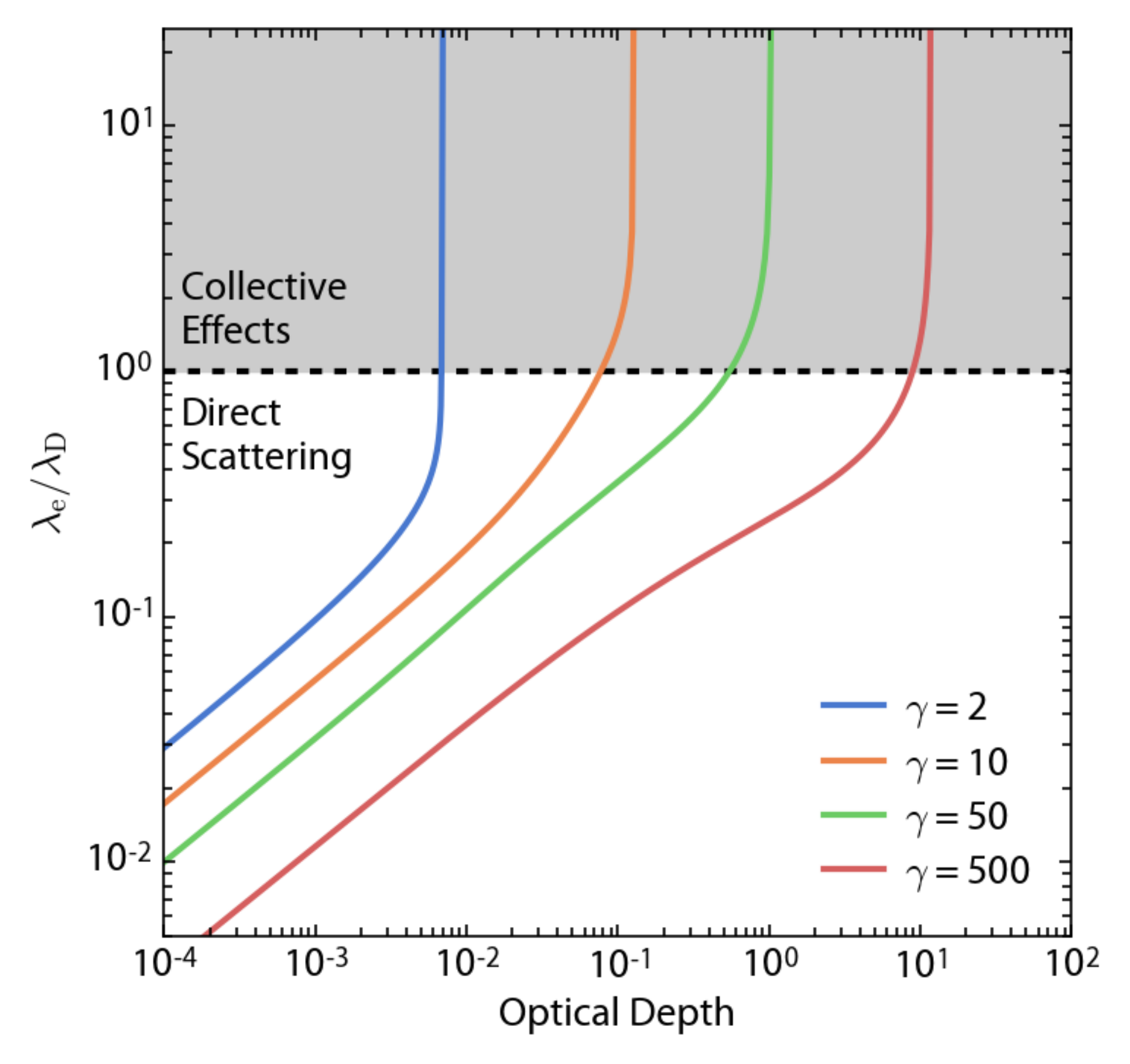}
\caption{Ratio of the de Broglie wavelength $\lambda_{\rm e}$ of the electron to the Debye length $\lambda_{\rm D}$ of the plasma as a function of optical depth for four different initial electron energies. In the outer regions of the atmosphere, the electron wavelength is much shorter than the Debye length, so individual scattering events dominate the energy loss. At higher optical depth, the electron wavelength is longer than the Debye length, so the dominant mode of energy loss is through the excitation of Langmuir waves.}
\label{fig:lam_e_lam_D}
\end{figure}  

We expect direct scattering or excitation of plasma modes to dominate the energy loss rate in different regions of the atmosphere. The relative importance of these effects is determined by the de Broglie wavelength of the electron or positron,
\begin{equation}
\lambda_e = \frac{h}{\gamma m_e \beta c}\;,
\label{eq:lambda_e}
\end{equation}
and the Debye length of the plasma,
\begin{equation}
\lambda_D = \sqrt{\frac{k_B T}{4 \pi n_e e^2}}\;,
\label{eq:lambda_D}
\end{equation}
where $k_B$ is the Boltzmann constant and $T$ is the temperature of the plasma.
In particular, where the de Broglie wavelength  of the impinging particle is much larger than the Debye length of the plasma, the charges in the plasma are screened and the particle is traveling through an effectively neutral material. In this case, individual scattering events become unimportant and the energy loss is dominated by the excitation of Langmuir waves. In the opposite regime, where the de Broglie wavelength is much smaller than the Debye length, the particle interacts with each electron individually and the scattering term dominates. Figure~\ref{fig:lam_e_lam_D} shows the ratio of the de Broglie wavelength of an impinging particle to the Debye length of the plasma as a function of the optical depth for several electron energies in the return current (the details of the atmospheric calculations are described in \S3). As this figure shows, direct scattering is expected to dominate in the outer parts of the atmosphere, whereas collective effects become important at optical depths $\gtrsim$ 0.1--1, depending on the energy of the return current.

\section{Radiative Transfer in the Neutron-Star Atmosphere} \label{section:equations}

To model the properties of the neutron star atmosphere, we follow the standard approach of \citet{mih78}. Our main goal here is to study the conditions under which the energy of the bombarding beam of particles is deposited above or below the photosphere and not to generate detailed models of the atmosphere or the emerging radiation. For this reason, we will assume a fully ionized hydrogen atmosphere, neglect the effects of electron scattering, and perform all our calculations in the gray limit, using Rosseland mean opacities. Additionally, we adopt a one-dimensional model and neglect the spatial distribution of the return current on the stellar surface. 

We begin with the zeroth moment of the equation of radiative transfer in the form
\begin{equation}
\frac{d H}{d z} = \chi_R(S - J). 
\label{eq:B-J}
\end{equation}
Here, $S$ is the source function, $J$ is the zeroth moment of the specific intensity, $H$ is the first moment of the intensity, and $z$ is the standard depth variable in a plane-parallel atmosphere and is measured from the 
top of the atmosphere. The opacity $\chi_R$ is the Rosseland mean opacity, 
\begin{equation}
\chi_R = \int_0^\infty\frac{1}{\chi_{\rm ff} + \chi_{\rm T}} \frac{\partial B_{\nu}}{\partial T} d \nu,
\label{eq:chi_Ross}
\end{equation}
where $B$ is the Planck function. Here $\chi_{\rm ff}$ and $\chi_{\rm T}$ are the free-free and Thomson opacities, respectively:
\begin{align}
\chi_{\rm ff} &= \frac{4 r_e^3 m_e^2 c^5}{3 h} \sqrt{\frac{2\pi}{3 k T m_e}} n_e^2 \nu^{-3} \left(1 - e^{\frac{-h \nu}{k T}}\right)g_{\rm ff} \label{eq:chi_ff},\\
\chi_{\rm T} &= \sigma_{\rm T} n_e, \label{eq:chi_T}
\end{align}
where $r_e$ is the classical electron radius, $T$ is the local temperature of the plasma, $g_{\rm ff}$ is the Gaunt factor for free-free emission, and $\sigma_{\rm T}$ is the Thomson cross-section. The Rosseland mean opacity also relates the physical depth $z$ to the optical depth $\tau$, i.e.,
\begin{equation}
\frac{d\tau}{dz} = \chi_R
\label{eq:dzdtau}
\end{equation}
For the free-free Gaunt factor, we use equation~(5) of \citet{gro78}. This approximation is obtained by performing a fit to the numerical results of \citet{kar61} and is accurate to within 15\% for wavelengths between 1 and 1000~\AA~\citep{gro78}. Our results are insensitive to variations of the Gaunt factor of this magnitude. 

The right hand side of equation~(\ref{eq:B-J}) represents the local imbalance between the heating and cooling terms, which must be equal to the specific rate of energy deposition from the return current. We designate this heating term as
\begin{equation}
Q^+ = -\chi_R(B - J).
\label{eq:Qplus_def}
\end{equation}
The specific rate of energy deposition, in turn, is related to the energy loss of the particles in the return current given by equations~(\ref{eq:ee}) and (\ref{eq:pe}) scaled by the density and velocity of the particles, i.e.,
\begin{equation}
Q^+ =  -n_{\rm RC} \beta c \frac{d E}{dz},
\label{eq:dzdt_heat}
\end{equation}
where $n_{\rm RC}$ is the number density of particles in the return current. 

To solve for the temperature and density profile in the bombarded atmosphere, we also introduce the first moment of the transfer equation, which gives us a relation between the first moment $H$ and the second moment $K$ of the intensity:
\begin{equation}
 \frac{d K}{d z} = \chi_R H.
\label{eq:dKdtau}
\end{equation}

Using equation~(\ref{eq:dKdtau}) and recognizing that, in local thermodynamic equilibrium, the source function $S$ is the Planck function $B$, we can rewrite equation~(\ref{eq:B-J}) as
\begin{equation}
 \frac{d^2 K}{d z^2} = \chi_R^2 (B - J).
\label{eq:d2Kdtau2}
\end{equation}

Lastly, we can relate the zeroth moment of the intensity $J$ to the second moment $K$ by the variable Eddington factor $f$:
\begin{equation}
J = \frac{K}{f}.
\label{eq:var_edd}
\end{equation}
We set $f=1/3$ throughout the atmosphere. This allows us to write the local conservation of energy in the form
\begin{equation}
\chi_B(B - \frac{K}{f}) + Q^+ = 0,
\label{eq:T}
\end{equation}
where we have defined the Planck weighted opacity as
\begin{equation}
\chi_B = \frac{\pi }{\sigma T^4}\int_0^\infty(\chi_{\rm ff} + \chi_{\rm T})B_{\nu} d\nu.
\label{eq:chi_Planck}
\end{equation}

The final equation we need is obtained by considering hydrostatic balance in the atmosphere, which yields a differential equation for the local electron density in the plasma given by 
\begin{equation}
 \frac{dn_e}{d z} = - \frac{n_e g m_p^2}{2 y_G^2 T},
\label{eq:dndtau}
\end{equation}
where $g$ is the local gravity, $m_p$ is the proton mass, and $y_G$ is the gravitational redshift.

We use a fourth-order Runge-Kutta algorithm to solve the coupled differential equations~(\ref{eq:ee}), (\ref{eq:d2Kdtau2}), and (\ref{eq:dndtau}), subject to the condition in (\ref{eq:T}), with $Q^+$ given by (\ref{eq:dzdt_heat}). At each point, we numerically find the temperature for which $\chi_B$, $\chi_R$, $B$,  and $Q^+$ satisfy equation~(\ref{eq:T}) and calculate the optical depth from equation~(\ref{eq:dzdtau}). At the outer edge of the atmosphere, we set the following boundary conditions:
\begin{align}
\tau_0 &= 10^{-6}, \label{eq:tau0}\\
H_0 &= \sigma T_{\rm eff}^4, \label{eq:H0}\\
K_0 &= \frac{1}{2}H_0, \label{eq:K0}\\
n_{e,0} &= 10^{17} {\rm cm^{-3}}, \label{eq:ne0} 
\end{align}
where $T_{\rm eff}$ is the effective temperature of the atmosphere. For the sample calculations shown below, we set $k T_{\rm eff}$ to 0.4~keV (which is typical for NICER sources) and $n_{\rm RC}$ such that the energy flux from the return current over the area of the hotspot is equal to the flux of thermal radiation at temperature $T_{\rm eff}$ over the same area. That is to say, 
\begin{equation}
n_{\rm RC} = \frac{\sigma T_{\rm eff}^4}{\pi m_e c^3 \beta (\gamma - 1)}.
\label{eq:n_rc}
\end{equation}

\section{Atmospheric Heating}

\begin{figure}
\includegraphics[width=3.5in]{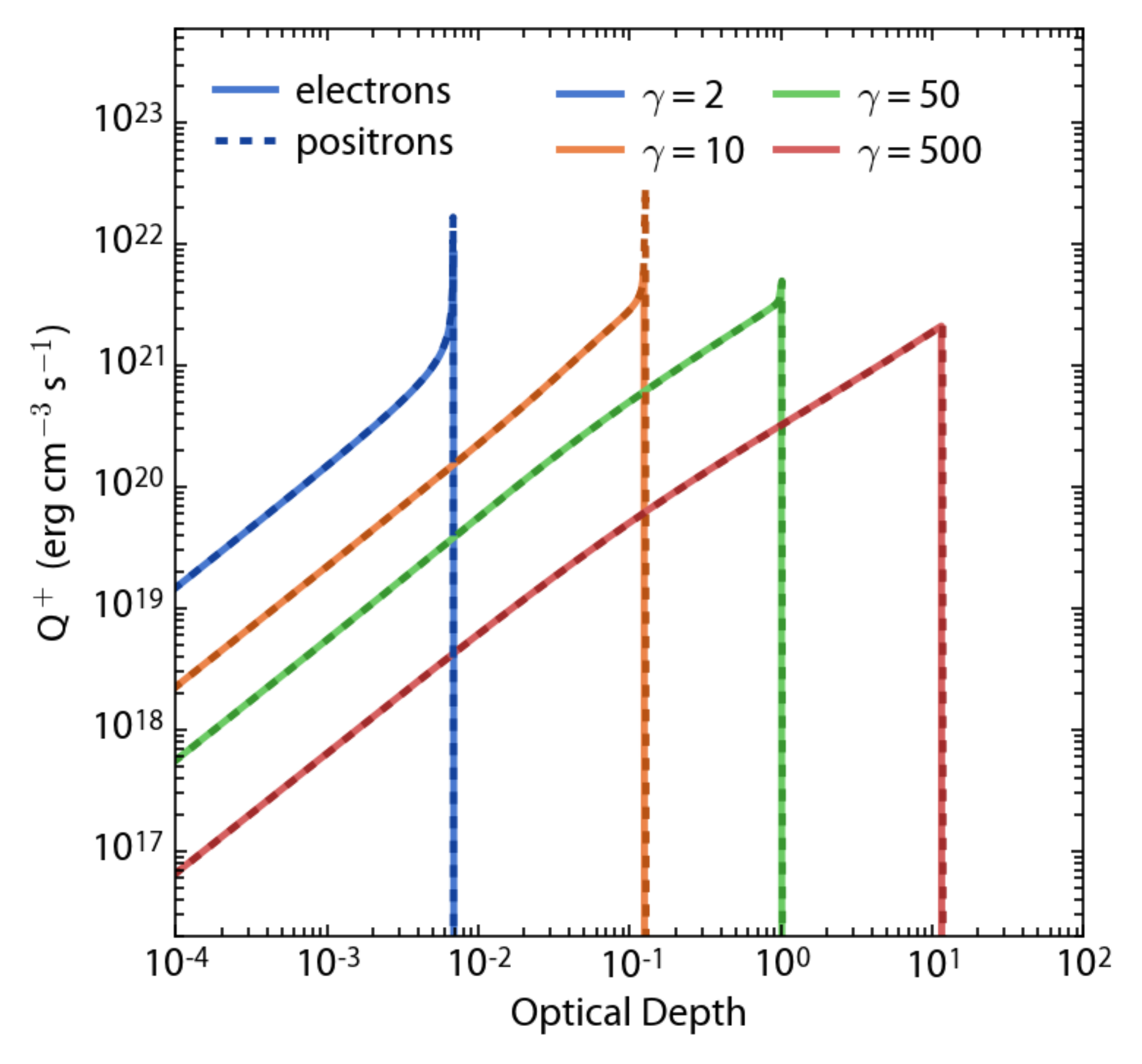}
\caption{Rate of energy loss as a function of optical depth for four different monoenergetic electron beams. For each energy, the solid line corresponds to a pure electron beam, while the dashed line corresponds to a pure positron beam. In all cases, the difference between the electron and positron cross-section (and hence the energy deposition rate) is negligible. }
\label{fig:Q_tau_fixed}
\end{figure}  

Figure~\ref{fig:Q_tau_fixed} shows the rate of energy deposition $Q^+$ as a function of the optical depth for three different initial particle energies. Although the scattering terms for electrons and positrons differ as described in \S\ref{section:Motivation} above, the rate of energy deposition is nearly identical for both species. In both cases, scattering is more efficient for less relativistic particles. Therefore, the rate of energy deposition increases as the energy of the particle decreases. This leads to a sharp peak in the energy deposition rate at the effective stopping depth of the particle.  

As expected, higher energy particles penetrate deeper into the neutron star atmosphere. Relatively low-energy particles ($\gamma \approx 2-10$) are effectively stopped before reaching $\tau = 1$. Much higher energy particles justify the deep-heating assumption, reaching a peak energy deposition rate at $\tau \gg 1$. 

\begin{figure}
\includegraphics[width=3.5in]{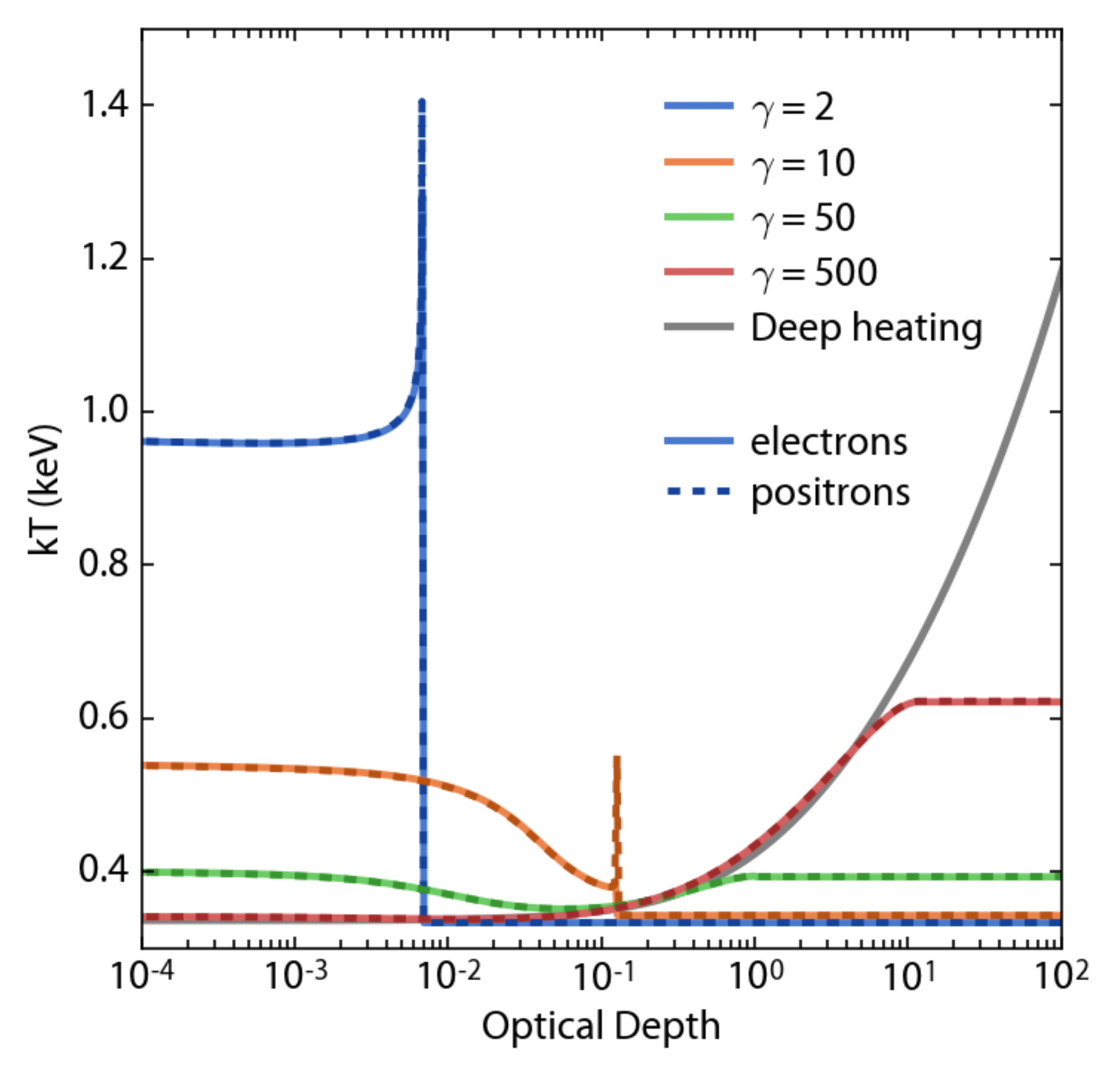}
\caption{Temperature profiles for monoenergetic return currents. Again, solid lines denote pure electron beams and dashed lines pure positron beams. The broad grey line indicates the analytic deep-heating solution which corresponds to a heating source at very large optical depth. At particle energies above $\gamma \sim 50$, energy deposition occurs at a large enough optical depth that the deep-heating solution is a reasonable approximation.}
\label{fig:t_tau_fixed_g}
\end{figure}

Figure~\ref{fig:t_tau_fixed_g} shows the temperature profiles in atmospheres bombarded with mono-energetic electron or positron beams with different energies. For high-energy particles, we recover the temperature profile of an atmosphere in radiative equilibrium. In this case, the energy deposited at low optical depth is negligible and the deep-heating approximation is valid. Lower energy particles, on the other hand, scatter more effectively and therefore heat the shallower regions of the atmosphere. We find that a temperature inversion forms, in which the outer layers of the atmosphere are hotter than the inner layers. 

For comparison, we also plot the analytic solution for a deep-heating atmosphere. In this case, which corresponds to an atmosphere in which all of the energy is deposited at infinite optical depth, the temperature profile is given by
\begin{equation}
T = T_{\rm eff}\left[\frac{3}{4} \left(\tau + \frac{2}{3}\right)\right]^{\frac{1}{4}}
\label{eq:deep_heat}
\end{equation}
\citep{mih78}. As shown in Figure~\ref{fig:t_tau_fixed_g}, this approximation is valid for high-energy return currents, but fails when $\gamma~\lesssim~50$. In this case, the deep-heating approximation is no longer valid and a different model for the atmosphere is needed. 

In our model, we have made the assumption that the atmosphere is thermalized at all optical depths and, therefore, the source function is a blackbody spectrum. This assumption is valid, if the heating timescale $t_h$ is much longer than the collisional timescale $t_c$, allowing the particles in the plasma to reach thermal equilibrium between successive heating events. 

The heating timescale depends on the ratio of the energy of the particles in the plasma to the heating rate $Q^+$, i.e., 
\begin{equation}
t_h = \frac{n_e k T}{Q^+}
\label{eq:t_h}
\end{equation}
The thermalization timescale is related to the collision timescale of particles in the plasma. Following \citet{spi62}, we write this timescale as
\begin{equation}
t_c = \frac{11.4\times10^6 A^{1/2} T^{3/2}}{n Z^4 \ln{\Lambda}},
\label{eq:t_c}
\end{equation}
where $A$ is the particle mass in units of the proton mass, $Z$ is the atomic number of the plasma species, and $\ln{\Lambda}$ is the Coulomb logarithm, which takes on values between $\approx 10-18$ for conditions considered here.

\begin{figure}
\includegraphics[width=3.5in]{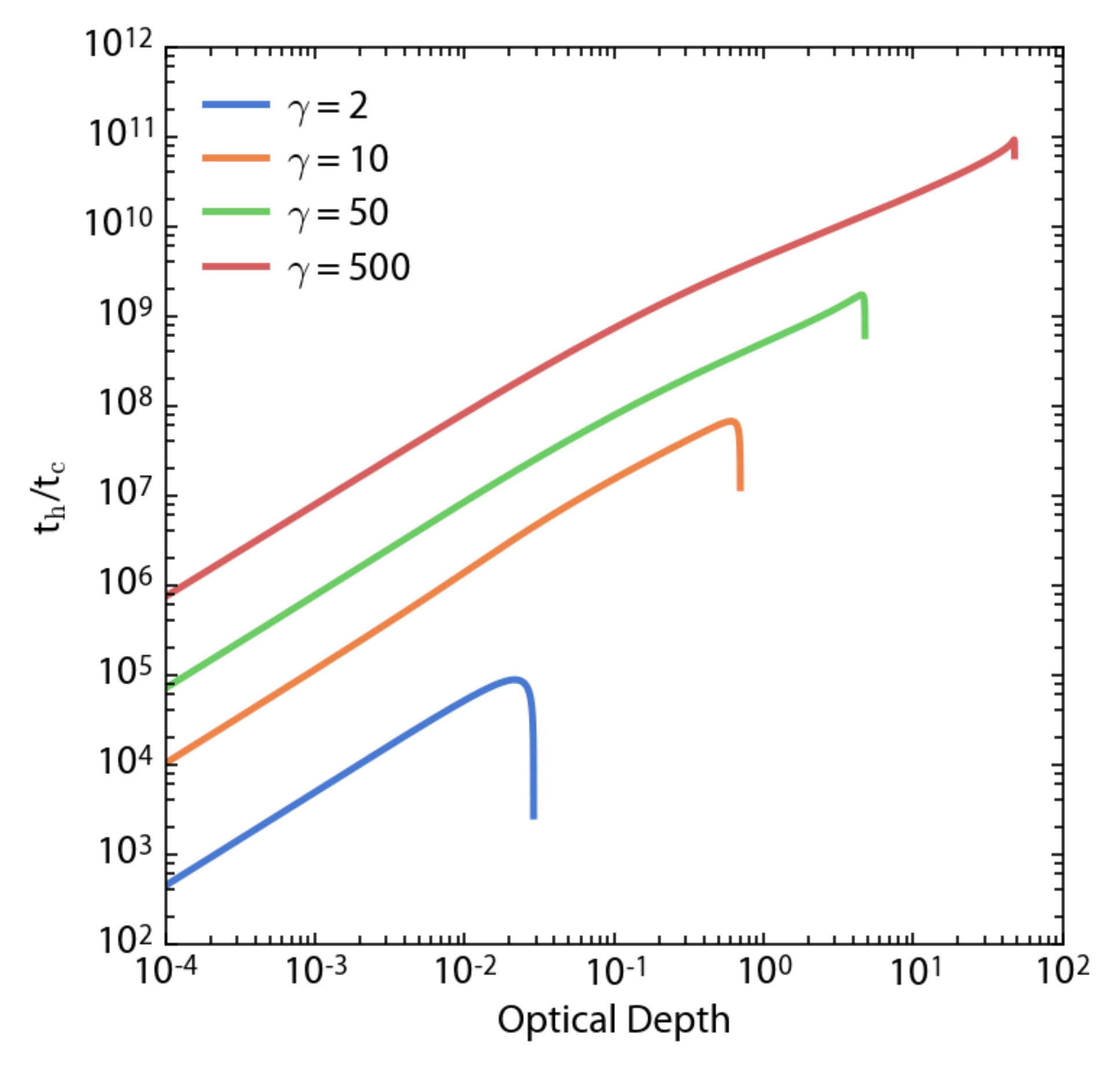}
\caption{Ratio of heating timescale to collisional timescale for four different atmospheric models. In all cases, the heating timescale is much longer than the collisional timescale, indicating that the atmosphere is in thermal equilibrium.}
\label{fig:t_heat_t_cool}
\end{figure}

Figure~\ref{fig:t_heat_t_cool} shows the ratio of the heating timescale to the thermalization timescale as a function of optical depth in several of our atmosphere models. Even in the outer regions of the atmosphere (where collisions are less frequent) and for a low return-current particle energy (which maximizes the heating efficiency), the particles in the atmosphere thermalize more than 100 times faster than they are heated. In deeper regions of the atmosphere and for higher particle energies, the ratio is even higher, justifying our assumption of thermal equilibrium.

\section{Return Current Energy Distributions}\label{dist}
In realistic situations, neutron-star return currents do not consist of mono-energetic beams of particles. Observations of gamma-ray pulsars as well as simulations of pulsar magnetospheres indicate that particles in return currents follow a power-law energy distribution up to very high Lorentz factors (e.g., \citealt{har01}, \citealt{cer16}, \citealt{bra18}). Although the gamma-ray emission is dominated by and constrains the highest energy particles, less energetic particles provide the largest contribution to the atmospheric heating. The particle distribution at these lower energies is less well known, but PIC simulations suggest the power-law distribution extends down to low Lorentz factors. We, therefore, adopt a power-law energy spectrum of electrons and positrons given by
\begin{equation}
N(\gamma) = N_0 \gamma^\alpha,
\label{eq:N_gamma}
\end{equation}
where $N_0$ is a normalization parameter and $\alpha$ the power-law slope. We set a minimum particle energy $\gamma_{\rm min}$ as the low-energy cutoff. 

The only modification this energy distribution introduces to the procedure described in \S4 is in calculating the local heating due to the return current particles. In the case of an extended energy distribution, we integrate over the particle energies to find the heating rate:
\begin{equation}
\frac{Q^+}{\chi} = -n_{\rm RC} \beta c \int_{\gamma_{\rm min}}^{\infty} \frac{d E_\gamma}{dz} d\gamma
\label{eq:Q+_gamma_dist}
\end{equation}
As before, we set the normalization $N_0$ such that the total particle energy flux on the surface is equal to the flux of a blackbody emitting at $T_{\rm eff}$.

\begin{figure}
\includegraphics[width=3.5in]{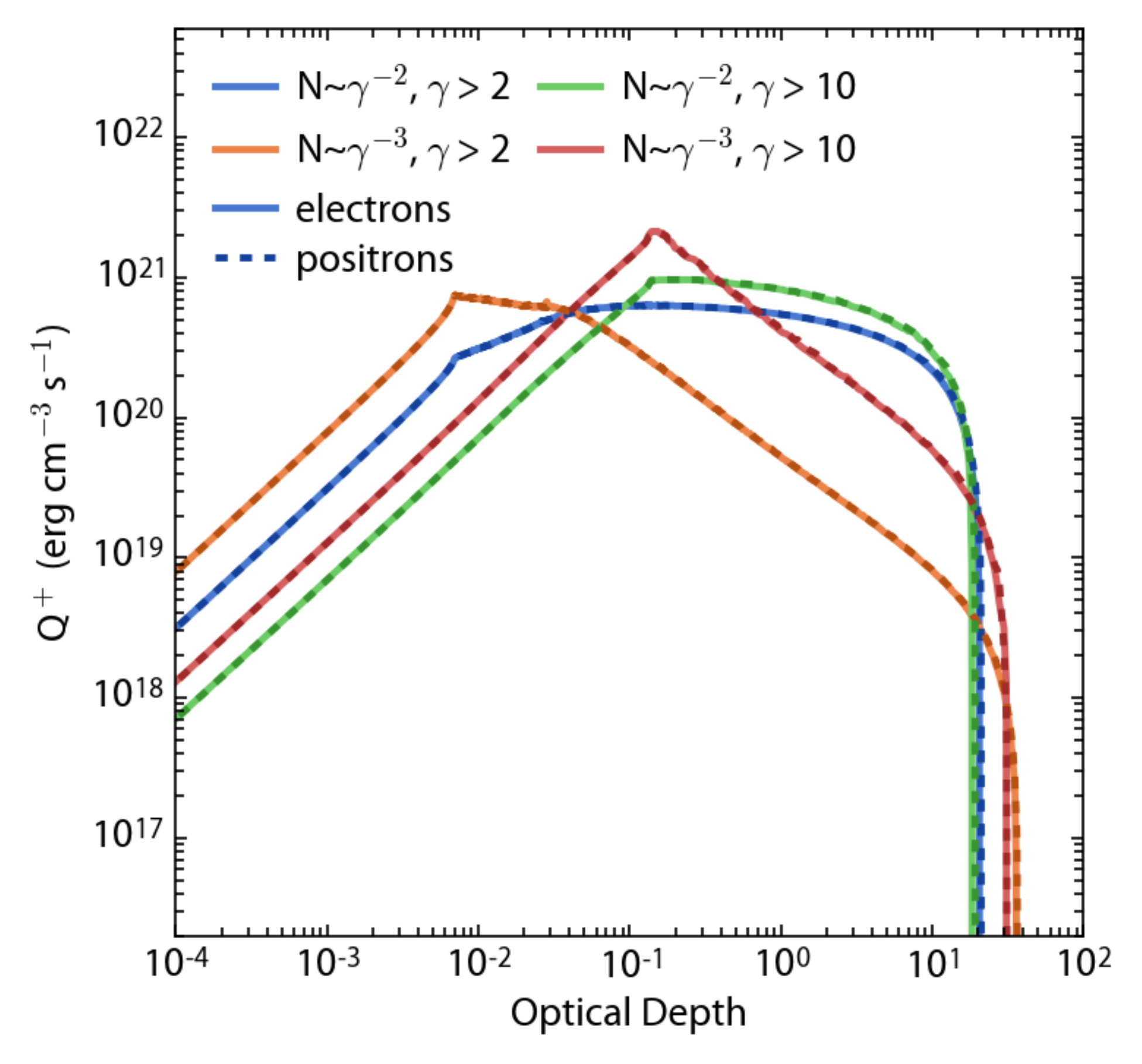}
\caption{Energy deposition rate for return currents with different energy distributions. The different curves correspond to two different power law indices as well as two different cutoff energies (see text). As before, solid lines denote electron beams while dashed lines denote positrons.}
\label{fig:Q_tau_g_dist}
\end{figure}

Figure~\ref{fig:Q_tau_g_dist} shows the deposited energy as a function of optical depth for two different power-law slopes $\alpha$ and two different cutoff energies $\gamma_{\rm min}$. As in Figure~\ref{fig:Q_tau_fixed}, we show the results for both pure electron and pure positron beams, although the difference in the heating produced by the different particle species is negligible.

\begin{figure}
\includegraphics[width=3.5in]{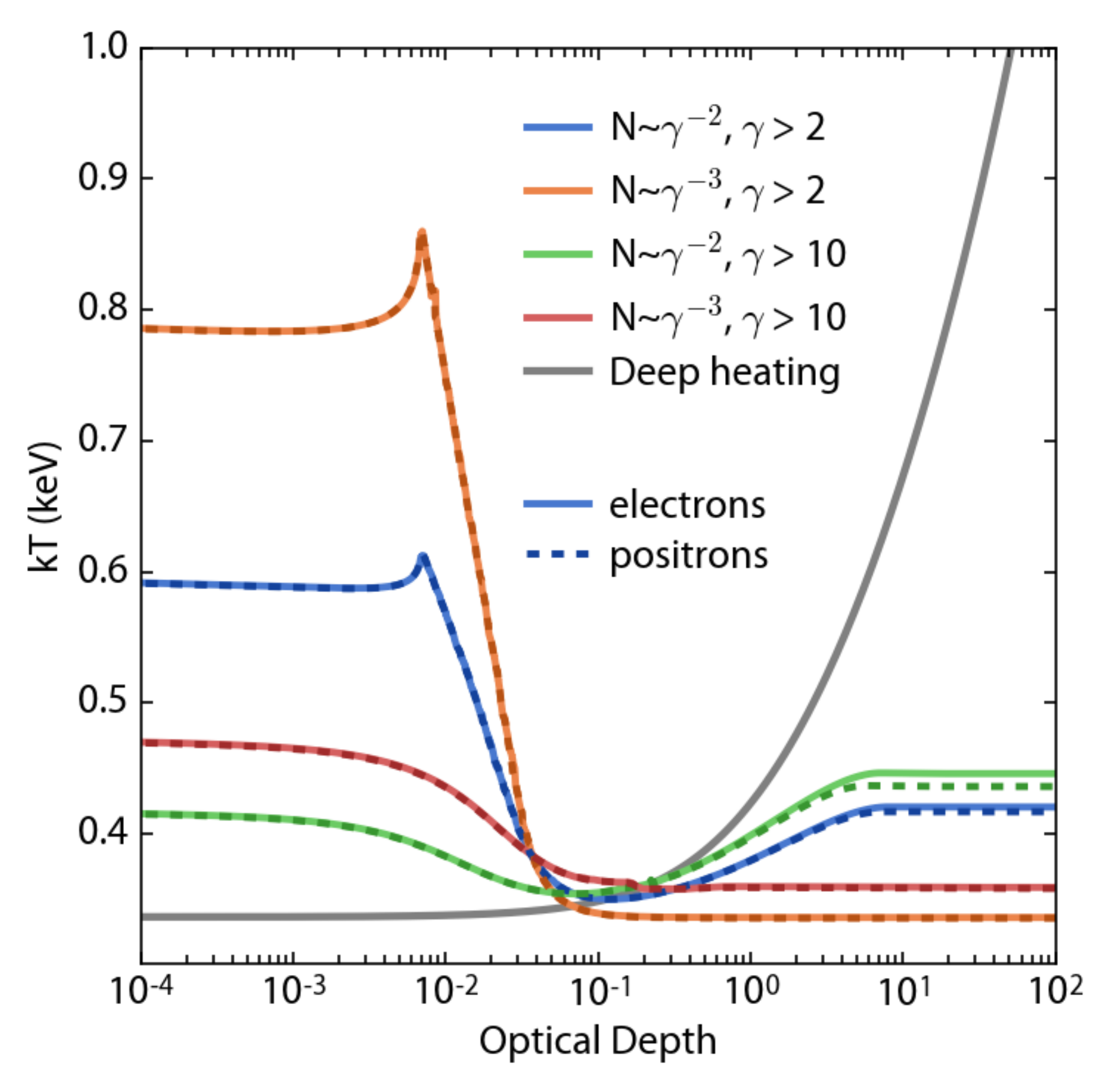}
\caption{Temperature profiles for different return-current energy distributions. For all the energy distributions considered here, the temperature profiles differ significantly from the deep-heating solution, shown as a broad grey line.}
\label{fig:t_tau_g_dist}
\end{figure}

Figure~\ref{fig:t_tau_g_dist} shows the temperature as a function of optical depth resulting from return currents with the same energy distributions as in Figure~\ref{fig:Q_tau_g_dist}. In all of these cases, the temperature distributions in the atmosphere differ significantly from the deep-heating model.

\section{Beaming of Emergent Radiation}

\begin{figure*}
\includegraphics[width=7in]{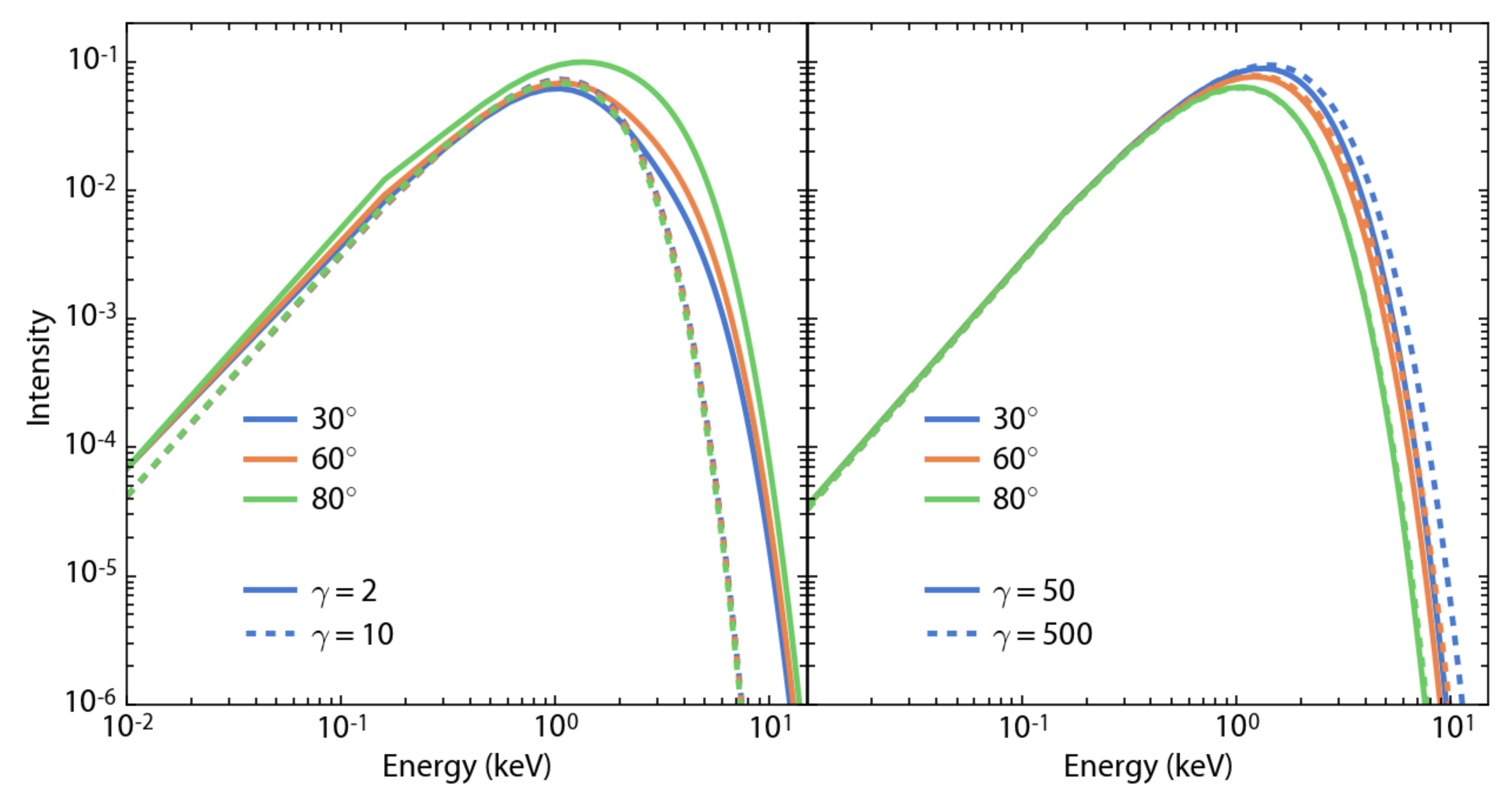}
\caption{Emergent spectra at different angles from the normal. The left panel shows spectra two lower energy models, while the right panel shows spectra for higher return current energies. The spectra are both broadened and shifted as a result of the atmospheric temperature profiles.}
\label{fig:spec_fixed}
\end{figure*}

\begin{figure*}
\includegraphics[width=7in]{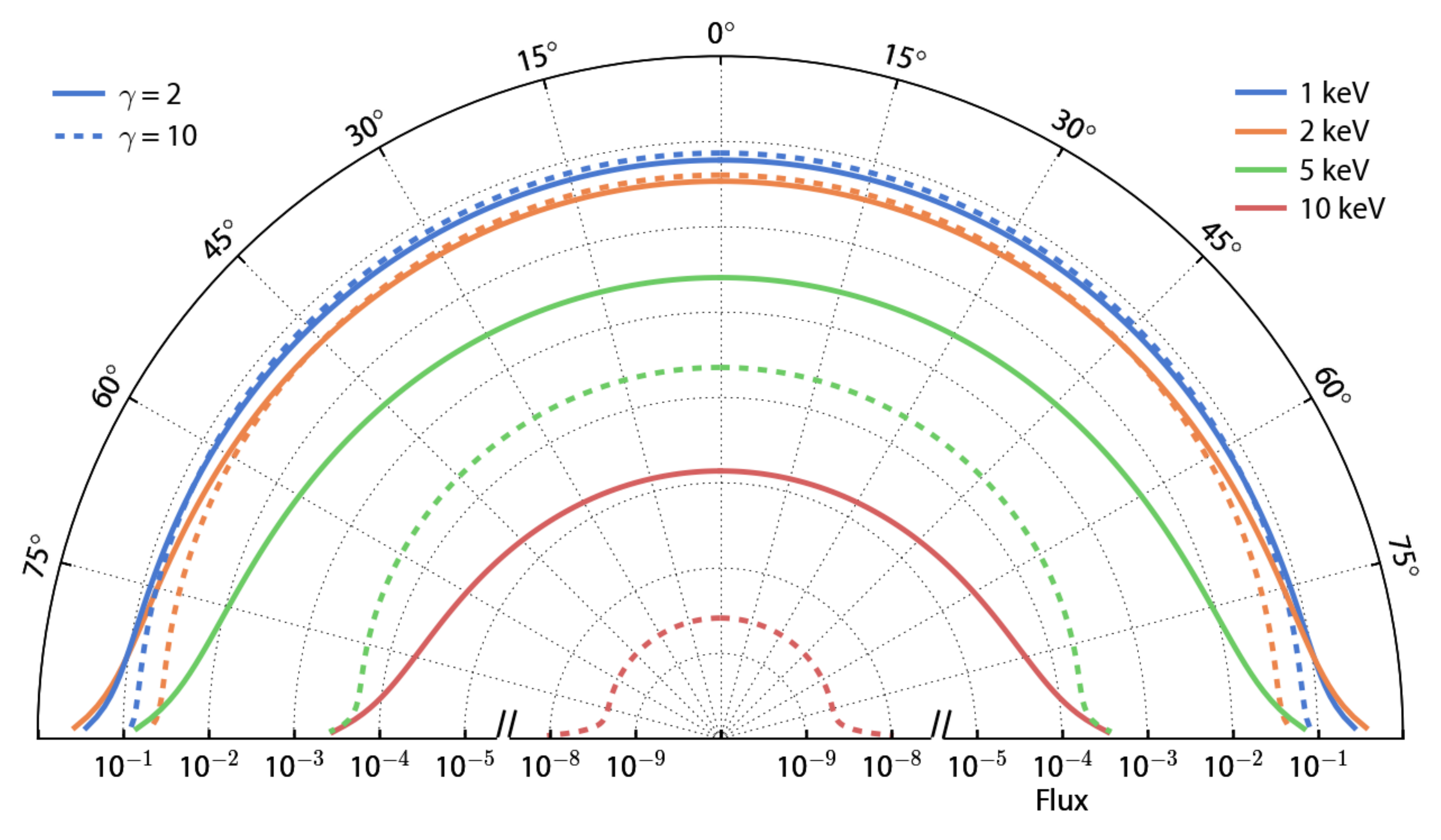}
\caption{Beaming functions for two atmosphere models with lower energy return currents at four different photon energies. Especially at the lowest return-current energies, there is significant limb brightening at all photon energies.}
\label{fig:beam_fixed_1}
\end{figure*}

\begin{figure*}
\includegraphics[width=7in]{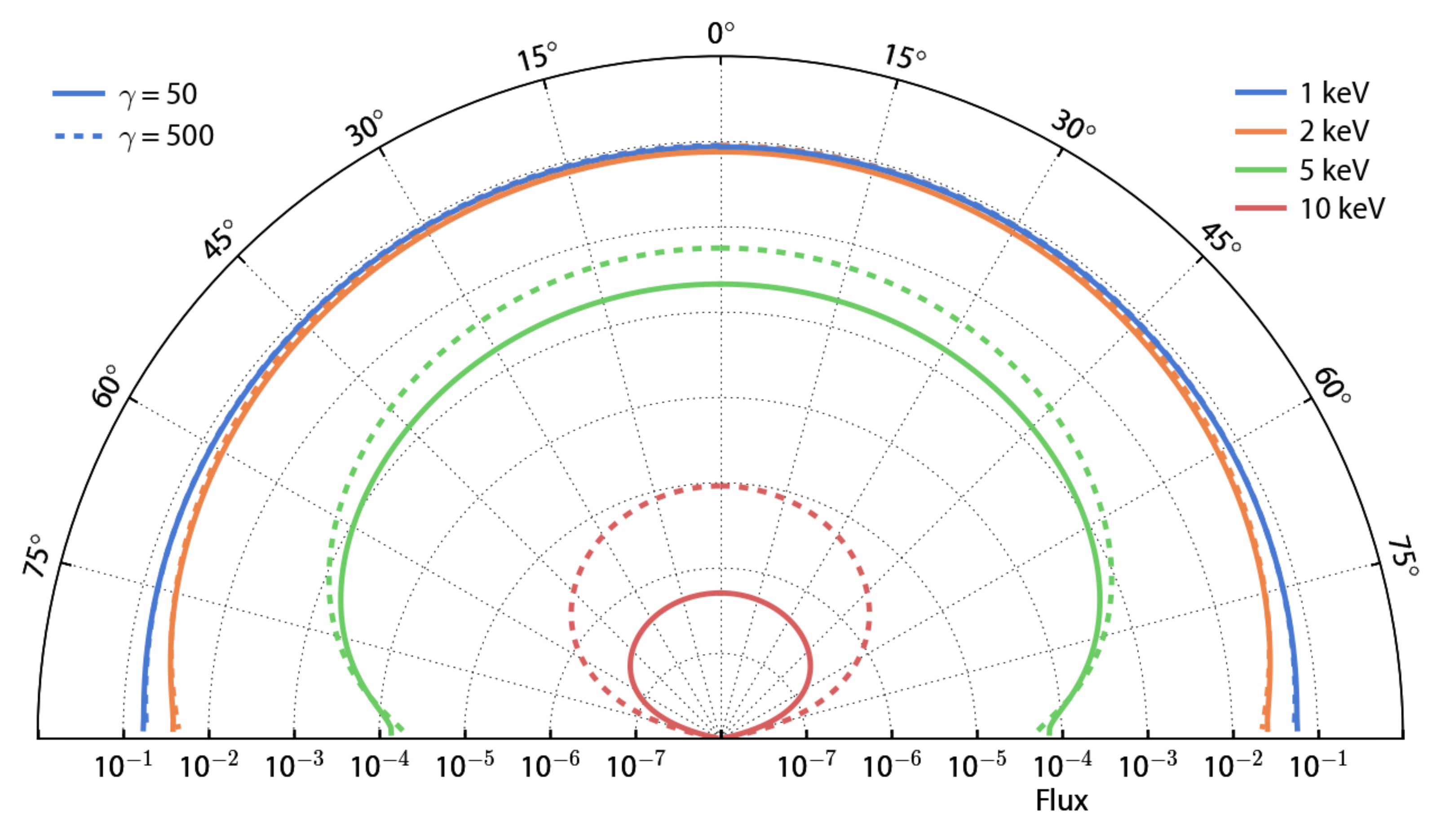}
\caption{Beaming functions for two atmosphere models with higher energy return currents. In this case, the emission is limb-darkened to varying degrees.}
\label{fig:beam_fixed_2}
\end{figure*}

\begin{figure*}
\includegraphics[width=7in]{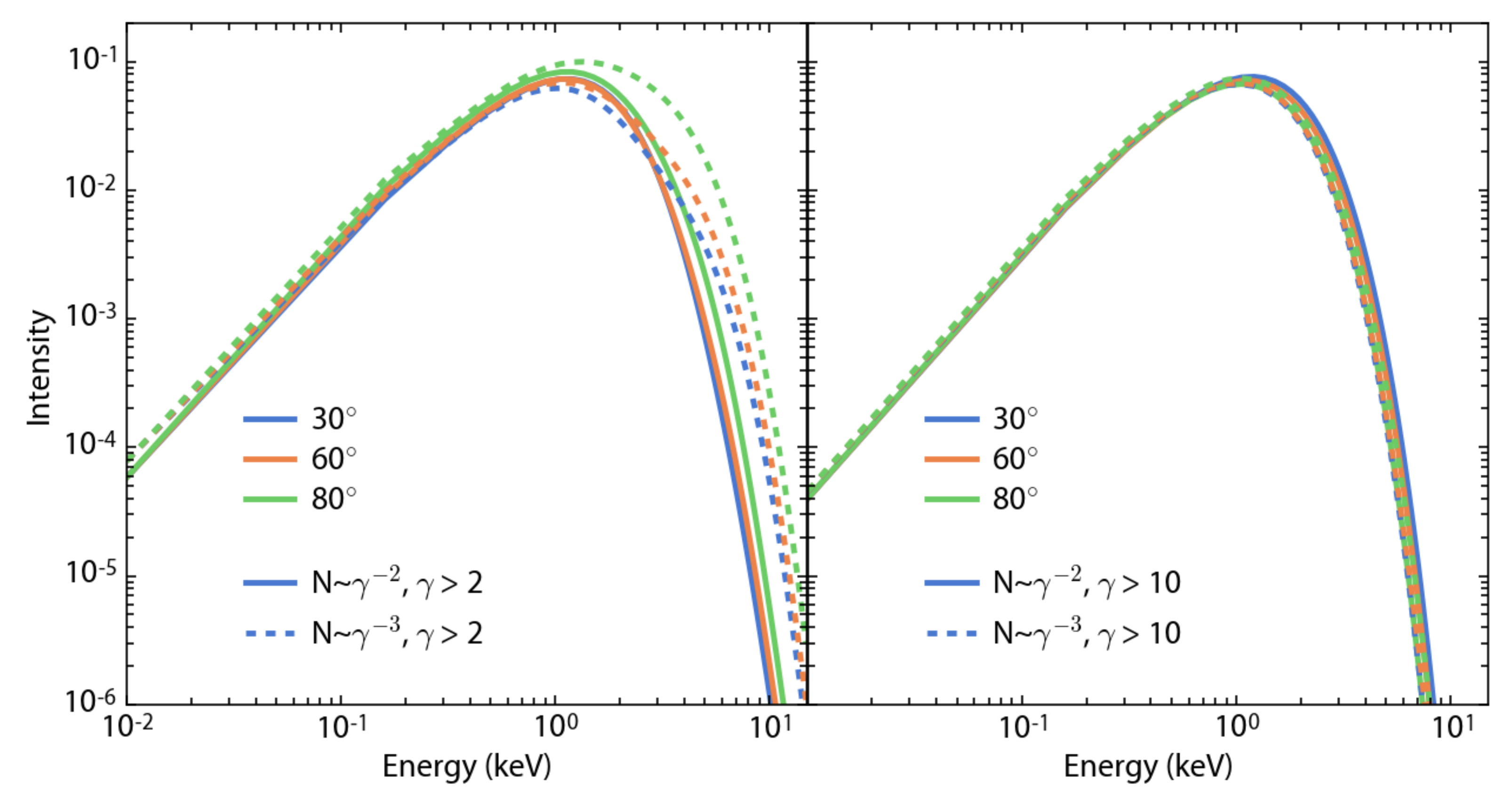}
\caption{As in Figure~\ref{fig:spec_fixed} but with return-current energy distributions as described in Section~\ref{dist}. Again, those distributions that include low-energy electrons display significant broadening and hardening depending on the viewing angle}
\label{fig:spec_dist}
\end{figure*} 

\begin{figure*}
\includegraphics[width=7in]{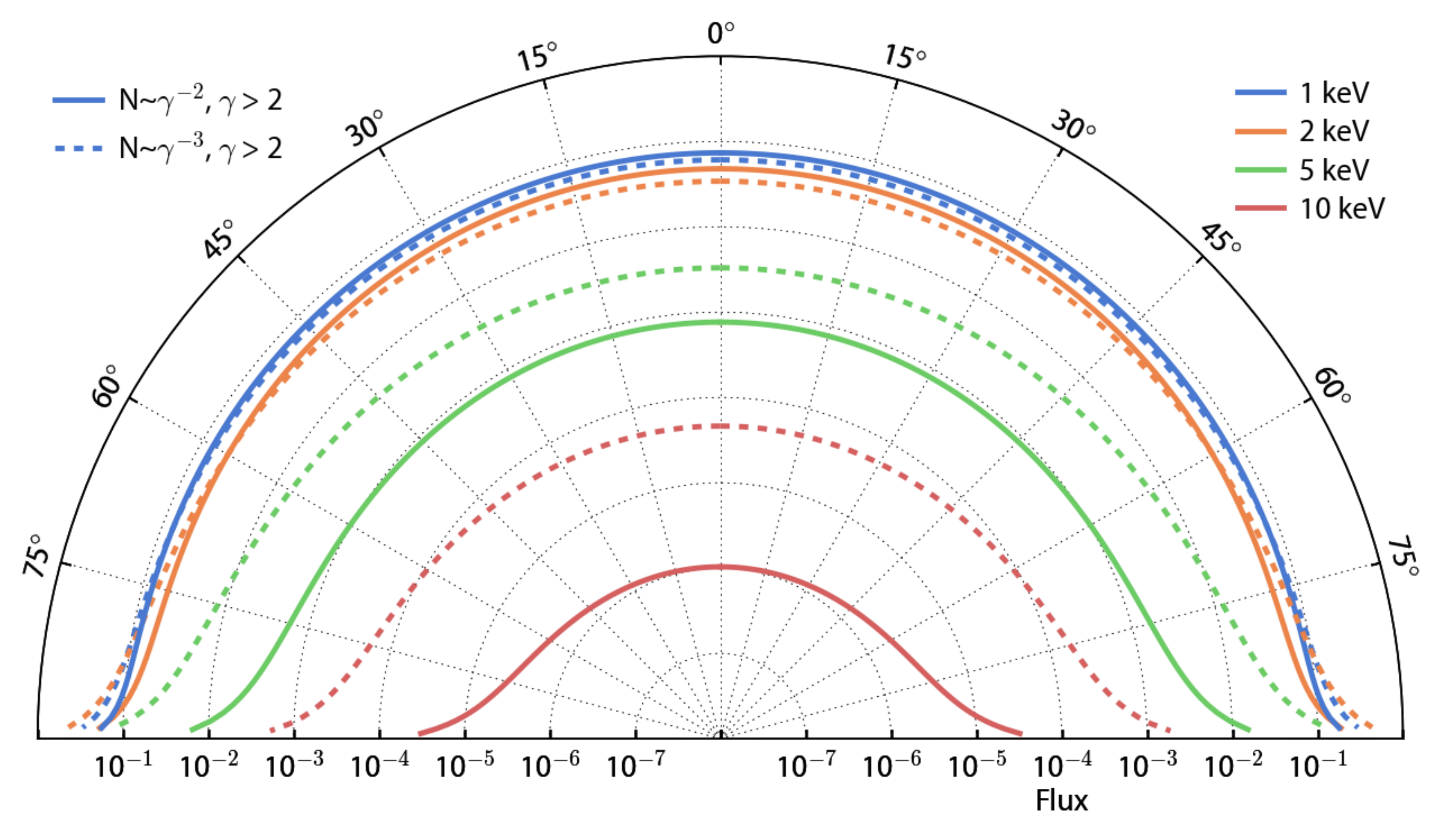}
\caption{Beaming functions for return-current energy distributions with a low-energy cutoff of $\gamma = 2$. As for the low-energy monoenergetic return current models, these atmospheres display significant limb brightening.}
\label{fig:beam_g_dist_10}
\end{figure*}

\begin{figure*}
\includegraphics[width=7in]{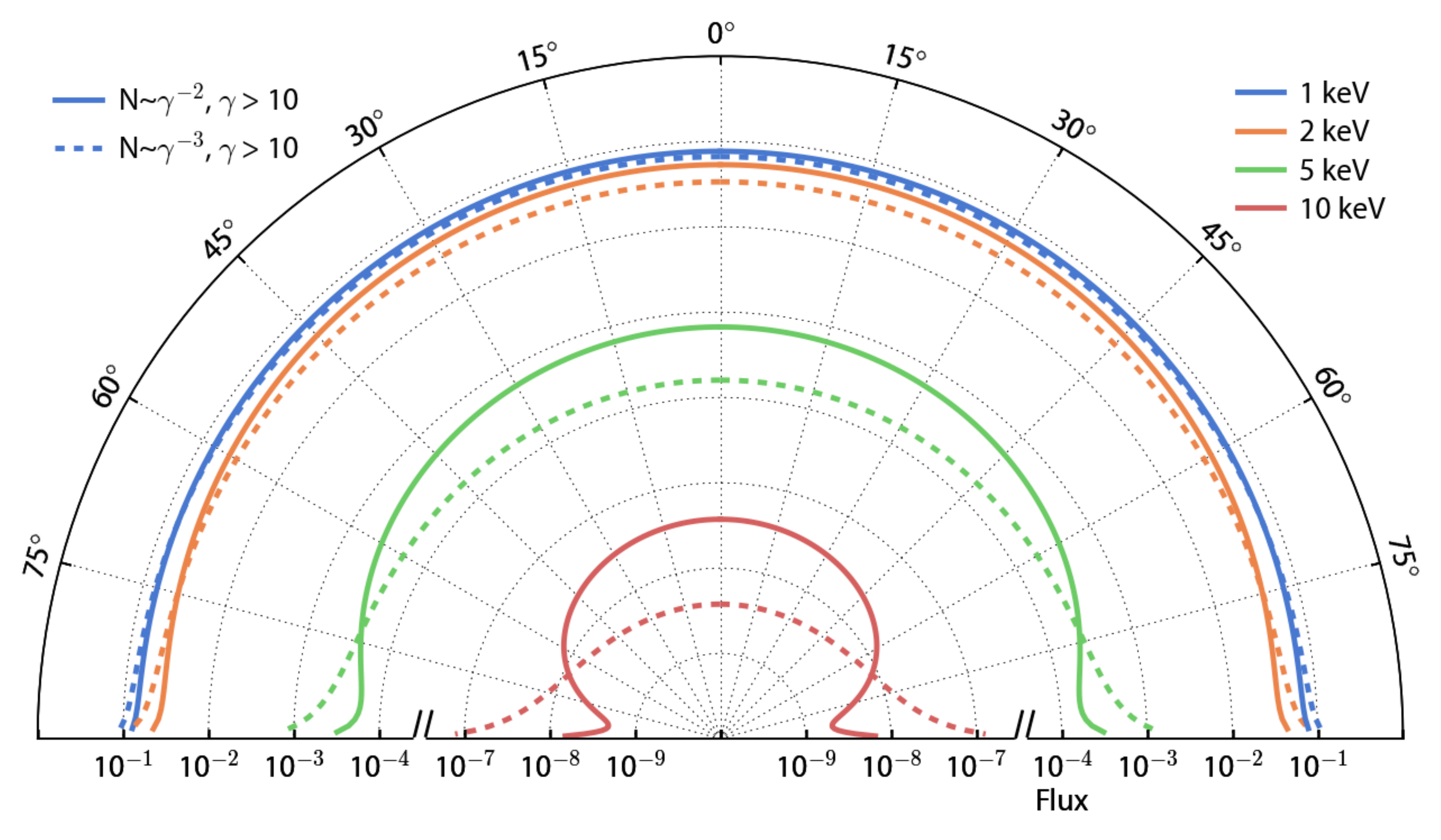}
\caption{Beaming functions for return-current energy distributions with a low-energy cutoff of $\gamma = 10$. These models show varying degrees of limb darkening, especially at higher photon energies.}
\label{fig:beam_g_dist_2}
\end{figure*}

The primary observable effect of return-current heating of the magnetic poles of neutron stars is the brightness oscillations that occur in X-ray lightcurves as these hotspots move into and out of view as the star spins. The observed pulse waveforms of these oscillations are strongly affected by the beaming of the emerging radiation, which, in turn, depends on the temperature profile of the atmosphere. Indeed, different temperature gradients and/or temperature inversions can lead to forward-peaked (pencil), isotropic, or cone (fan) emission patterns from the hotspot, which can significantly alter the X-ray pulse waveform.

In order to calculate the beaming function for a given atmosphere model, we make use of the radiative transfer equation,
\begin{equation}
\frac{d I_\nu}{d s} = \chi (- I_\nu + S_\nu).
\label{eq:beam_rad_trans}
\end{equation}
Because the atmosphere is in local thermodynamic equilibrium and we have neglected the effects of electron scattering, the source function $S_\nu$ is the Planck function $B_\nu$. At an angle $\theta$ from the surface normal, the differential optical depth is given by $d\tau_\nu = \chi_\nu ds = \chi_\nu dz/\cos(\theta)$, and we can integrate equation~(\ref{eq:beam_rad_trans}) to find the emergent intensity, i.e., 
\begin{equation}
I(\theta) = \int_{0}^{\infty} B_\nu(T) e^{-\tau} d\tau_\nu.
\label{eq:beam_rad_trans_2}
\end{equation}
The calculations described in Section~\ref{section:equations} fully specify the opacity and temperature at each depth $z$ in the atmosphere for a given heating function. We can, therefore, numerically integrate equation~(\ref{eq:beam_rad_trans_2}) from the surface ($\tau = 0$) to large optical depth (which we take to be $\tau = 1000$ for numerical purposes) for each photon energy and incident angle to find the emergent spectrum and beaming function.


Figure~\ref{fig:spec_fixed} shows the spectrum of radiation at three different angles of incidence for a range of return current energies. Figures~\ref{fig:beam_fixed_1} and~\ref{fig:beam_fixed_2} show the corresponding beaming functions for several photon energies in the NICER bandpass. As expected, models with lower return-current electron energies that lead to temperature inversions in the neutron-star atmosphere result in fan-beam patterns and harder spectra at high incidence angles. Those models with high return-current electron energies and corresponding deep-heating temperature profiles result in limb-darkened emission and harder spectra at viewing angles close to the surface normal. 



Figure~\ref{fig:spec_dist} shows the emergent spectra for the four different return-current energy distributions discussed in Section~\ref{dist}. Figures~\ref{fig:beam_g_dist_10} and~\ref{fig:beam_g_dist_2} show the corresponding beaming functions. As before, the atmosphere models in which the heating from return current particles occurs principally at large optical depth show a decreasing intensity at larger angles, whereas models with shallow heating are brighter at large angles to the normal. 

\section{Discussion}

Our numerical models of the atmospheric structure of pulsar polar caps
demonstrate that the resulting spectra and beaming patterns of the
emerging radiation depend rather strongly on the energy spectrum of
the bombarding beam of particles. This is especially true if
particles with Lorenz factors $\gamma\lesssim 50$ carry a significant
amount of the energy content of the beam.

The shape of the energy spectrum of magnetospheric particles in the
low-$\gamma$ regime is very difficult to infer observationally or
simulate numerically. Observationally, the best constraints on the spectrum
of magnetospheric currents come from modeling the $\gamma$-ray
properties of the pulsars. However, this part of the photon spectrum
is determined primarily by particles that have substantially higher
Lorenz factors ($\gamma\gg 100$). In numerical Particle-In-Cell
simulations of magnetospheric currents, the energy spectrum of the
magnetospheric particles shows a significant component of low-$\gamma$
particles, and its details are determined by the rate of injection of
charges and the inclination of the magnetic axis with respect to the
rotation axis~\citep[see, e.g.,][]{cer16,kal18}. However, because of
numerical limitations related to resolving effects at the plasma
frequency, the parameters of the simulations are chosen such that the
maximum Lorentz factor achieved is small ($\gamma\lesssim 1000$). In
the absence of other prior information on the energy spectra of the
particle beams, modeling of the atmospheric properties of polar caps
in order to measure neutron-star radii using NICER observations will
require a parametric description of the particle energy spectra and an
investigation of the dependence of the results on the various
parameters.

Calculating pulse profiles for the beaming patterns of the bombarded
atmospheres is beyond the scope of this paper and will be explored
elsewhere. However, we can estimate the impact of the anisotropic
beaming on the pulse profiles using the semi-analytic estimates
discussed in \cite{pou06} and \cite{oze16b}. If we write the angular
dependence of radiation emerging from the stellar surface as
\begin{equation}
I(\theta)=I_0 (1+h \cos\theta)\;,  
\end{equation}
where $h$ is a parameter that measures the degree of anisotropy and depends
on photon energy, then the fractional amplitude of the pulse profile scales
as (see eqs.~[20] and [22] of \cite{oze16b})
\begin{equation}
  r1=\frac{(1+2 h q)v}{q+h(q^2+v^2/2)}\;.
  \label{eq:r1}
\end{equation}
The two auxiliary parameters $q$ and $v$ depend on the
neutron-star compactness
\begin{equation}
  u\equiv \frac{2GM}{Rc^2}\;,
\end{equation}
the inclination $i$ of the observer, and the inclination $\theta_{\rm
  B}$ of the magnetic axis with respect to the rotational axis via the
relations
\begin{equation}
  q\equiv u+(1-u)\cos i\cos\theta_{\rm B}
\end{equation}
and
\begin{equation}
  v\equiv (1-u)\sin i\sin\theta_{\rm B}\;.
\end{equation}
Equation~(\ref{eq:r1}) can be solved analytically for the neutron-star
compactness, given a measurement of the amplitude $r_1$ of the pulse
profile. The solution is complicated algebraically but we can study its
behavior by simplifying it as
\begin{equation}
  u\simeq -\frac{h}{2}+r_1 + {\cal O}(h^2,r_1^2, hr_1)\;,
  \label{eq:sol}
\end{equation}
where we have expanded the solution to first order in $h$ and $r_1$
and evaluated it at $i=\theta_{\rm B}=\pi/2$. Equation~(\ref{eq:sol})
shows that, assuming that all other parameters in the system are
known, the accuracy with which the compactness of a neutron star can
be measured via pulse profile modeling is determined at nearly equal
parts by the accuracy of the measurement of the amplitude of the pulse
profile and by the accuracy of the prior knowledge of the beaming of
radiation.

A second effect of the presence of anisotropy in the beaming of the
emerging radiation is the change in the harmonic content of the
pulse profile. Indeed, the ratio of the amplitude $c_2$ of the second
harmonic to the amplitude $c_1$ of the fundamental becomes
(see~\citealt{pou06}, eq.[50])
\begin{equation}
  \left(\frac{c_2}{c_1}\right)_{\rm aniso}\simeq h \sin i \sin\theta_{\rm B}.
\end{equation}
If an unsuitable beaming profile is used to model an observed pulse
profile, then the ratio of the amplitudes of the harmonics will
be attributed to rotational effects that have a similar dependence, i.e.,
\begin{equation}
  \left(\frac{c_2}{c_1}\right)_{\rm rot}\simeq \left(\frac{4\pi f R}{c}\right)
  \sin i \sin\theta_{\rm B}\;,
\end{equation}
where $f$ is the neutron-star spin frequency~\citep{psa14a}, and lead to
a biased measurement of the neutron-star radius.

These estimates are based on the assumption of a small, circular hotspot. Our calculations of the atmosphere above are one-dimensional and therefore do not take into account the spatial distribution of return-current particles on the stellar surface. Several analytic (e.g.\ \citealt{gra17}) and numerical (e.g.\ \citealt{phi2018}) calculations of the distribution of the return current on the stellar surface indicate that the resulting hotspot may deviate significantly from a circular shape and have a complex temperature distribution. These complications are also expected to affect the resulting pulse profiles. A full analysis of predicted lightcurves from non-uniform hotspots is beyond the scope of this work and will be discussed elsewhere. 

In light of these results and the estimates of their potential impact on X-ray pulse profiles, it will be important to incorporate the effects of shallow heating due to particle bombardment in the atmosphere when modeling the high signal-to-noise data obtained from rotation-powered pulsars with NICER.

\acknowledgments
We  thank A. Harding, A. Timokhin, A. Bret, and J. Dexter for helpful discussions. This work was supported by NASA grant NNX16AC56G.

\bibliographystyle{yahapj}
\bibliography{bibliography}

\end{document}